# BUSINESS PROCESS MANAGEMENT SYSTEMS IN PORT PROCESSES: A SYSTEMATIC LITERATURE REVIEW

Alicia Martín-Navarro, María Paula Lechuga Sancho*, and José Aurelio Medina-Garrido

*Department of Organizational Business, INDESS. Universidad de Cádiz*
*Corresponding author

This is the version submitted to the International Journal of Agile Systems and Management (IJASM). The final published version can be found at https://doi.org/10.1504/ijasm.2020.109245

We acknowledge that Inderscience holds the copyright of the final version of this work.

Please, cite this paper in this way: Martín-Navarro A, Sancho Lechuga MP, Medina-Garrido JA (2020) Business process management systems in port processes : a systematic literature review. International Journal of Agile Systems and Management, 13(3):258–278.

## Abstract

Business Process Management Systems (BPMS) represent a technology that automates business processes, connecting users to their tasks. There are many business processes within the port activity that can be improved through the use of more efficient technologies and BPMS in particular, which can help to coordinate and automate critical processes such as cargo manifests, customs declaration the management of scales, or dangerous goods, traditionally supported by EDI technologies. These technologies could be integrated with BPMS, modernizing port logistics management. The aim of this work is to demonstrate, through a systematic analysis of the literature, the state of the art in BPMS research in the port industry. For this, a systematic review of the literature of the last ten years was carried out. The works generated by the search were subsequently analysed and filtered. After the investigation, it is discovered that the relationship between BPMS and the port sector is practically non-existent which represents an important gap to be covered and a future line of research.

## 1. Introduction

Increasingly, commercial ports are gaining momentum as logistics centres (Ruiz-Torres et al., 2018). In fact, they are important links in the logistics chain in international trade and the growth of foreign trade (Fioresi De Sousa et al., 2019). Increasing volumes of international trade demand more efficient cargo handling worldwide. For this reason, the port logistics chain is attracting greater attention from both professionals and researchers. These chains are more complex nowadays and new configurations of shipping services are needed (Ramírez-Nafarrate et al., 2017). In this context, not only is greater infrastructure needed to support dock operations, but further digitization of processes is essential to increase efficiency and competitiveness in port management. Traditionally, in ports, inflexible information systems have been used which did not allow the correct coordination of all processes, continuous improvement, or even the integration with other software tools.

This type of system has among its main drawbacks the difficulty of eliminating useless information avoiding redundancies, the low capacity to store data so that it is available when the user deems it convenient, problems both in providing security by avoiding the loss of information, as in the intrusion of unauthorized personnel or even not generate useful output information for users making the decision-making process far more difficult. That is why the interactivity and flexibility of the information systems constitute a key point in the success or failure of organizations management (Bran, 2015). The need of greater flexibility in information systems is related to the current trend of modern organizations to focus on processes and therefore, focus on clients instead of paying attention to hierarchical and functional structures (Reijers, 2006). In this sense, processes cross departmental boundaries and organizations become more horizontal. A process-oriented company usually applies the concept of Business Process Management (hereinafter BPM) (Kohlbacher, 2010). BPM is a management discipline focused on aligning all aspects of an organization, based on an analysis, control and maintenance of the business processes. Specifically, BPM aims for the organization to gain in efficiency and effectiveness and its objective is for the company to continuously improve its processes (Malaurent and Avison, 2016). This management philosophy generated the need to implement specific information systems oriented to process management. Initially, specific solutions were programmed for particular processes. These systems, called workflows, were rigid and inflexible. In fact, any change in the process required rescheduling the workflow. However, the continuous improvement of business processes advocated by BPM demanded more dynamic and flexible software. To this end, companies introduce the automation of their processes through business process management systems (BPMS), which are defined as a set of software methods, techniques and tools that support the design, control and analysis of business processes (Aalst *et al.*, 2003). A BPMS provides tools to model and graphically redesign processes, without dwelling on database design, coding nor even connections with other systems. BPMS do not replace existing systems, but rather integrate the use of available applications, the tasks that make up each process, and the people who perform each task. The workers may follow at each moment the instructions given by the graphic design of the processes that are carried out with the BPMS, pointing out to each worker when they have to carry out the task, what task they have to do, what it consists of, and what information, software and resources they have in order to accomplish the task (Arjonilla Domínguez and Medina Garrido, 2013).

The literature on the study of BPMS has increased in recent years, which shows a growing interest in research in this field (Martín, 2017). BPMS has also been published in numerous journals of different disciplines. For this reason, and according to (Işik *et al.*, 2013) and (Roeser and Kern, 2015), it is clear that BPMS is a multidisciplinary paradigm. Given the interest in the academic literature, and the increasing use of BPMS in different port terminals, the objective of this work is to study, through a systematic literature review, the state of the art in research on BPMS and its application in terminals or port processes in the international area. In this regard, the research question for this study would be: What are the main works that study the port processes in which BPMS is applied? From the selected papers it will be analyzed the following: (1) Number of publications per year, (2) contributions by country, (3) contributions by author, (4) contributions by area of knowledge, (5) type of study, (6) research topic, and (7) type of software tool.

Therefore, the work is structured as follows. In section 2, a review is made on BPMS and port management and their trends. Section 3 describes the methodology used. In section 4, the main results are presented and analyzed. And finally, in section 5, the work is concluded and the main contributions and lines of future action are provided.

## 2. Background

### 2.1. Business Process Management Systems

Business processes management (BPM) has evolved as an important research paradigm, reaching considerable maturity. It provides proven methods that can conquer current and future challenges (vom Brocke *et al.*, 2014) and companies have embarked on initiatives to manage business processes to gain competitive advantage (Castro and Erazo, 2009). Business processes management is the art and science of supervising how work is done to ensure solid results and take advantage of opportunities for improvement (Dumas et al., 2013). Tucek (2015) defines business processes management as a methodology to evaluate, analyze and improve key processes, depending on the needs and desires of customers.

At present, much attention is paid to the efficiency of the processes (Amadi-echendu and Kruger, 2015), due to the business environment that has become more competitive. Companies have had to reinvent themselves and work through networks, distributed elsewhere (Friedman, 2005). On the other hand, because process management breaks interdepartmental barriers, the business process can navigate between different people from different functional areas, for which information technologies are very useful. The creation of value in the processes aims to increase the efficiency and effectiveness of the process, automating and rearranging tasks, and creating interconnected chains of processes (Davenport and Short, 1990; Hammer and Champy, 1993).

The last technological phenomenon in the market that allows for change is known as BPMS (Wong, 2013). The existing technology was not flexible, once installed, to adapt to new needs. However, thanks to the BPMS, the processes become more executable, and the work can be automated without the need to program. This allows the systems to adapt to the company and not the other way around. Consequently, the business process is released from its mold. And it is through an agile administration of processes that the entire value chain can be monitored, improved and continuously optimized (Smith and Fingar, 2003).

The BPMS need to design a model of the process that is used to obtain the graphic representation of a current or future process within the organization.[1] Analysis of the business is a prerequisite for the design and development of information systems, since it can reduce the amount of problems caused by misunderstandings between people who work in the business area and information technology equipment (Bernardino Araújo et al., 2016). These systems are not intended to replace existing applications within the organization; they add those applications to a new process and use their information, integrating all that information in a more flexible and adaptable way than if software was reprogrammed (as is the case with workflows, which are also process automation tools). In this way, businesses become more agile (Wong, 2013). This greatly facilitates integration with other heterogeneous systems and, normally, serves as a platform for the development of process-based applications (Rhee et al., 2007). For AuraPoral[2], a software provider company, BPMS is the consolidated and unstoppable trend that is changing forever the way to manage the operations of companies and any organization in the world, allowing much greater flexibility, much greater automation and much greater power.

### 2.2. Port management

Globalization and the entry of more nations into the World Trade Organization has boosted the growth of maritime trade, causing the proliferation of increasingly polyvalent ports and allowing the intensification of transshipment activities, which has led to the need for specific services, as well as more dilated infrastructures (Feng et al., 2011). The ports, defined as places of contact between the different areas of circulation of goods and services, become spaces of convergence between transport systems and are integrated into a system of distribution of goods that requires more complex logistical developments. In this context, port performance is important because, due to its large volume and lower cost compared to other transports, it attracts 90% of global freight transport and international trade (Metalla et al., 2015), acting as an economic catalyst for income and employment. In addition, port efficiency is relevant to the competitiveness of countries, since their terminals are key elements of the international trade supply chain. All this without forgetting that the increase in environmental awareness has promoted the demand for transport by ship since the transport by water consumes less fuel than other modes of transport (Feng et al., 2011). That is why an inefficient port sector harms the insertion of a country in international trade by not allowing the competitiveness of its exports or making imports of products and basic inputs more expensive (Britto *et al.*, 2015).

At present, maritime ports base their competitiveness on technology and innovation in their logistics and supply chain processes (De Martino et al., 2015). The port of Singapore and the port of Hamburg are two outstanding examples of the relevance of ICTs and innovation, combined essentials that make the Asian and German seaports efficient and flexible world-class transshipment centers. The use of smart technologies can have a direct impact on the redesign of business processes in many seaports (Ferretti and Schiavone, 2016). Thus, the exchange of information in real time between cargo lines and port customs could greatly reduce the time of control and the transition of goods. Similarly, the monitoring of intermodality processes can lead to better operational efficiency.

---

[1] IIBA (2009), A guide to the business analysis body of knowledge (BABOK® guide) Version 2.0. 1st ed. International Institute of Business Analysis (IIBA), Whitby, ON.

[2] Available at: https://www.auraportal.com/es/producto/que-es-bpm/ (consulted 20/11/2017)

Bearing in mind that the services provided by the port terminals are determined by agreements between the companies that own the containers and the back-port terminals, whose main services are the shipment of these containers to the storage areas in their facilities where, in turn, they undergo inspection, repair, cleaning and testing processes until they are available to exporting companies. One way to create competitive advantage and improve these processes is by using a BPMS. This system allows processes to be more efficient and effective, while at the same time increasing productivity, improving profitability ratios and creating value for customers (Andrade et al., 2015).

According to Andrade et al. (2015) and Ferretti & Schiavone (2016), there are many business processes within the port activity that can improve thanks to the use of more efficient technologies and especially BPMS, which can help coordinate and automate critical processes, such as the cargo manifests, the customs declaration (DUA), the management of scales, or the management of dangerous goods, traditionally supported by EDI technologies (Yeo et al., 2011). These technologies could be integrated with BPMS, modernizing port logistics management by contributing directly to activities such as: the analysis of activities; the flow of documents and information in international trade procedures; the identification and prioritization of problematic areas that cause delays in the transfer of goods from the seller to the buyer; and the design of improvement measures to address these problem areas (for example, the simplification of processes and data, and the elimination of redundancies). In the same way, according to the Business Process Analysis Guide to Simplify Trade Procedure (2012)[3], this tool could contribute to the reduction of commercial procedures (including commercial, transport, regulatory and financial procedures); the simplification of documentary requirements and their alignment with international standards; and the automation of international commercial transactions and their associated electronic documents for one-stop systems and paperless commercial systems.

### 2.3. Main trends

Since BPMS are easily integrated with the use of any other pre-existing information system, they favour the incorporation of current new technological trends into process management. In this context, there are many initiatives around the transformation of ports into intelligent ports and the automation of logistics at European level. These new trends seek to respond to the challenges of optimisation of port space, port operation times, transport costs and productivity levels, maximization of loading and unloading flows challenges, physical limitations in the process, consumption of natural resources and reduction of environmental impact (Molina Serrano *et al.*, 2018).

Industry 4.0 is undoubtedly the new paradigm that is permeating the technological management of organizations and the management of ports. Although there is no consensus on the definition of the concept of industry 4.0, the common factor is that this change in the technological paradigm has to do with the digitalization and connectivity of all the technologies of the organizations (Hitpass and Astudillo, 2019). This concept goes beyond simple robotization or automation of production processes and services. It includes aspects such as IoT (Internet of things, where sensors, wearables, autonomous machines and even vehicles communicate among themselves and with users), sensorization, the interconnection of mobile devices that allow massive data collection, the intensive use of these data in production processes, Big Data analysis techniques and business intelligence, remote monitoring of processes and online decision-making, or the virtualization of processes to analyze multiple scenarios (Ferretti and Schiavone, 2016). Precisely, BPMS enhance process optimization by virtually representing them in different scenarios. Once implemented, BPMS allow the integration of the rest of the technologies of the industry 4.0 in processes management. The connectivity between all technologies favours vertical and horizontal integration of systems and processes. This allows the development of integrated value chains between different departments, or even between different organizations, so that they work in a cohesive way (Hitpass and Astudillo, 2019).

---

[3] Business Process Analysis Guide to Simplify Trade Procedures (2012). Ed. ESCAP. Available at: (consulted 22/11/17).

Digitization is pushing the maritime industry beyond its traditional boundaries and provides new opportunities to improve productivity, efficiency, performance and logistics sustainability (Kravchenko, A. V., & Makarenko, 2017). The concept of intelligent ports is postulated as a future trend that aims to adopt information technologies that allow intra- and inter-organizational transactions. This is particularly important to better coordinate cargo delivery and land-based operations, where different processes are interrelated (Heilig et al., 2017). The success of the digital transformation also lies in the adaptation of organizational processes, following the idea that digitalization is a means and not an end. In this effort, technologies such as BPMS emerge to systematically address process change and achieve better results at all levels, from operations to management. Companies seek to simplify their processes and try to minimize their dependence on rigid technologies. For this reason, the flexibility of BPMS allows companies to be more agile and efficient in achieving their goals (Paschek, Ivascu and Draghici, 2018).

Nowadays, several ports that use BPMS to automate their processes have been detected at a national and international level. These include the Port of Barcelona[4] or the Port of Rio Grande in Brazil. Similarly, an international multicontainer[5] company using BPMS to automate recurrent processes has been located and works with various ports in Latin America, such as Argentina, Peru, Panama, Paraguay and Uruguay, wholesaling in Ecuador, Colombia and Chile and interacting with strategic partners in Brazil, the United States and Europe.

## 3. Methodology

According to Alvesson & Sandberg (2011), what is currently known as systematic literature review offers a general overview, a synthesis and a critical assessment of previous research, and, as a consequence, the main difficulties related to existing knowledge can be extracted and new research problems identified. Thus, this section includes an exhaustive review of the scientific production that relates BPM systems with port terminals. Specifically, following the process suggested by Kitchenham & Charters (2007), a review protocol has been developed, fundamentally in five stages: (1) definition of the research question, (2) design of the search strategy, (3) selection of works, (4) data extraction and (5) data synthesis.

---

[4] Available at: http://www.portdebarcelona.cat/ (consulted 14/07/2017).
[5] Available at: https://www.flokzu.com/blog/es/clientes/clientes-multicontainer/ (consulted 20/11/2017).

To carry out this systematic study of the literature, the main indexes to be analyzed will be:

1. Number of publications per year (Tarhan et al., 2016).

2. Contributions by country (Rooney et al., 2014).

3. Contributions by author (Idri et al., 2016).

4. Contributions by area of knowledge.

5. Type of study (Wen et al., 2012).

6. Research topic.

7. Type of software tool.

### 3.1. Definition of the Research Question

Taking into account all the above, and following Tarhan et al., (2016) methodological approach, the first step consists on giving answer to the aforementioned question or research question (RQ):

   RQ: What are the main works that study the port processes in which BPMS is applied?

### 3.2. Search strategy design

For the definition of the search terms used in the databases, the main concepts referred to in the research question were used, including their synonyms and abbreviations. In addition to the term BPMS, the term workflow was included. Workflows represent a type of information systems that even though they are not the same as BPMS they are used for the same need, to automate the management of processes, although with a very different management philosophy. There are even authors who have come to use the terms BPMS and workflow interchangeably (Pistol and Bucea-Manea-Tonis, 2015). Finally, keywords previously used in relevant scientific articles in the field were included.

 Thus, in order to answer the research question and obtain as many articles as possible, the search was done by using keywords combined in pairs. The word-pairing was divided into the following two groups: BPMS, "Business Process Management", workflow, "Process modeling", "Information system" and "Information technology", alluding to BPMS, and port, "retroportuary terminal", container, "marine transport", export and "port logistics process", with respect to ports.

### 3.3. Literary resources

To search for the most relevant articles related to the research topic, the following six digital databases were used: EBSCO, Emerald Insight, Google Scholar, ISI Web of Science, Proquest,

Scopus. This search procedure is widely accepted and has been used in other systematic literature reviews within the area of information systems (Tarhan et al., 2016).

### 3.4. Search process

In order to find out the state in which the use of BPM systems for management in the port area is in terms of research, it has been decided to focus this study on existing scientific works up to the year 2017 and in English (Ramírez Correa and García Cruz, 2005; Echeverri and Cruz, 2014), in the social science databases previously mentioned and included in the virtual library of the University of Cádiz. And primary studies and non-reviews were searched, like Guinea et al. (2016).

### 3.5. Search results through the five databases

The search of articles and the application of the filters specified above generated the identification of 1653 works that referred to the topics BPMS and ports. In Figure 1, you can see the results obtained by electronic resource.

### 3.6. Selection of works

#### 3.6.1. Inclusion and exclusion criteria

In a systematic literature review, we must determine the inclusion and exclusion criteria that are rigorously applied for the selection of relevant literature, with minimum judgment on the part of the researchers. That is why these studies should involve several researchers and reach an agreement between them (Jalali and Wohlin, 2012).

The main exclusion criteria were established, leading to the rejection of all those works (1) of less than four pages, (2) not available in full text on the Internet (Rocha *et al.*, 2013), (3) not related to the subject of this research, which is management by business processes and ports and, (4) that treated the management by processes but did not make reference to the software tool. The selection of articles was done manually, reading the abstract. The items that were subtracted amounted to a total of 23 works, to which a quality evaluation was applied.

#### 3.6.2. Quality criteria

Quality assessment is necessary to delimit biases, provide a more accurate idea of potential comparisons and guide the interpretation of results (Turner et al., 2010). To assess the primary studies following Kitchenham & Brereton (2013) and Guinea et al. (2016), a generic set of questions is used to evaluate rigor, credibility and relevance, and thus be able to make the last screen to obtain the final sample. This quality instrument was developed by Dybå & Dingsøyr (2008) in its systematic literature review on software engineering, and is applicable to most studies.

Specifically, the quality of the study was evaluated by classifying it according to eight different criteria listed in Table I. Each of the questions had three optional answers: "Yes", "partially" and "no". Following Wen et al. (2012), these responses were scored as follows: "Yes" = 1, "partially" = 0.5, and "no" = 0. The quality evaluation of the articles is calculated by adding the scores of the responses to the previously defined questions.

**[INSERT TABLE I. Quality Criteria. ABOUT HERE]**

To ensure the reliability of the results of this review, the only results considered to be acceptable quality were those that had scores greater than 4 (50% of the perfect score) for the subsequent extraction and synthesis of data (Guinea et al., 2016). In Figure 1, the entire systematic literature review process is observed, showing the number of definitive works.

**[INSERT FIGURE 1. Phases of the systematic review of literature process and number of definitive articles.**

**ABOUT HERE]**

## 4. Results and discussion

The analysis of the data was carried out for the 22 works (see appendix) after the application of all the steps established in the literature on bibliographic reviews was obtained, depending on a series of indexes.

### 4.1. Number of publications per year

While not many papers have been found that relate BPMS to port processes, the first work was located in 2004. As can be seen in Figure 2, from that year the number of publications is gradually increased, and it is in 2013 when it has its highest peak. This increase may be the result of the strategic orientations on transport that were set out in the White Paper on transport published in 2001 and 2011. In this sense, in the last White Paper on transport, the one of 2011[6], the European Commission identified digitisation as one of the essential principles and objectives for achieving a single European transport area, contributing to improving its performance. Similarly, in other markets, such as Asia, Chinese ports are also driving automation and digitisation. One example is the world's largest automated container terminal, the Deepwater Port of Yangshan (Shanghai). Another example is Ningbo-Zhoushan (China), which has upgraded its equipment in order to get a more competitive shipyard.

In the same way, in 2013 there is an increase in academic interest in BPMS in ports because in the context of a major economic crisis, these systems allowed greater control over business processes, great support for critical decision making, great agility and rapid adaptation to changes via restructuring and optimizating processes and greater strategic orientation to achieve business objectives (Martín-Navarro et al., 2018). In this context, BPMS allows greater control over processes.

---

[6] COM(2011) 144 final, de 28 de marzo de 2011.

**[INSERT FIGURE 2. Number of publications per year. ABOUT HERE]**

*4.2. Contributions by country*

Among the countries that have literature that relates BPMS to port activities, China is the most productive country with four works, followed by South Korea that provides three publications. France and Italy have two and the rest of the countries only produce one. This can be seen in Table II.

The fact that main works come from Asian countries is in line with the results published in The Liner Shipping Connectivity Index (LSCI)[7] published by the United Nations Conference on Trade and Development (UNCTAD). The launch of this index, which measures container port connectivity, prompted ports in 2006 to invest in digitization and infrastructure and hinterland connectivity to expand their container service networks. More specifically, the UNCTAD report details seven measures to help improve the efficiency of working vessels while in ports and to help ensure the smooth delivery of containers inland. These measures include digitizing processes, modernizing ports and monitoring the connectivity of national, regional and global networks. The port with the highest score in the LSCI is Shanghai, followed by Singapore, Hong Kong and Busan in South Korea.

**[INSERT TABLE II. Contributions by country. ABOUT HERE]**

*4.3. Contributions by author*

This section contains the names of those authors who are most actively involved in the research on BPM tools in port activities. A total of 64 authors participated in the production of the 22 selected works. Most of them only contribute once but others do it more often. In this way, the most productive authors are B.N. Yahya and H. Bae with three publications each. Those authors with two works are R.I. Fadi, Y. Guo, L. Huang and Y. Wang. It is quite significant that 90,62% of the authors have only published one article, that only 6,25% have two publications and just 2 authors have three papers (this is the 3,12%). In short, there are no large producers in the field but rather small producers encompassing the state of the art.

*4.4. Contributions by area of knowledge*

Table III shows the most prolific areas of knowledge. More specifically, the works can be framed in five different areas. The technological areas are those that generate the greatest number of publications. In this context, the area with more weight is engineering, with more than 36% of works produced, eight issues exactly. With the same proportion, the 18% of the papers are written by authors in the Computer field, Information Systems, and Business Organization.

---

[7] UNCTAD's liner shipping connectivity index (LSCI) for 2019. Available at: https://unctadstat.unctad.org/wds/TableViewer/tableView.aspx?ReportId=92. Consulted in: 21st of january 2020

Finally, only two works were found in the logistics area. Not surprisingly, the most productive area of knowledge is engineering, since information and communication technologies are widely studied in the literature from both a technical and a managerial perspective. It is precisely in the engineering field, together with the Computer and Information System fields, that the design, development and integration of the BPMS are worked on, while the Business Administration and Logistics fields focus more on aspects of management and continuous process improvement.

**[INSERT TABLE III. Contributions by area of knowledge. ABOUT HERE]**

### 4.5. Type of study

Academic research is basically divided into two main areas, empirical research and theoretical research (Bhosale and Kant, 2016). Within the empirical research, which is that based on evidence, the works found have been classified into case studies and experimental studies. As for theoretical research, all theoretical papers have been included, regardless of whether they are literature reviews or conceptual papers. As it occurs in BPMS research from a management perspective (Martín-Navarro et al., 2018), no quantitative empirical works analysing BPMS in ports have been found. A plausible explanation is that the nature of the object of study, the ports, makes access to large samples very difficult. Of the works analysed, most are case studies, with a percentage that exceeds 63%. Specifically, 14 works have been found with this methodology. On the other hand, only one experimental study has been found. In regard to the theoretical works, seven papers were found, which represents 31.82% of total production. Table IV shows the classification by type of study.

**[INSERT TABLE IV. Type of study. ABOUT HERE]**

### 4.6. Research topics

Business process management is encompassed in the life cycle stages of the BPMS. These stages include the analysis, modelling, implementation, execution, control and optimisation of the processes (Minonne and Turner, 2012). Considering these stages, the selected works are classified in three different research topics, all of them referring to BPMS. Two of them refer to the life cycle, which are implementation and modelling, to which a third topic has been added, which is software architecture. As shown in Figure 3, the most researched topic is the modeling of processes through a BPMS, with a percentage of work of 68% ([2][3][4][5][6][7][8][10][12][16][17][18][19][20][22]). The papers remaining study the implementation of the tool (23%) and, from a design perspective, 9% of them analyses the architecture of the BPM software. Particularly, articles number [9], [13], [15] and [21] study implementation, and articles number [1] and [14] study architecture.

**[INSERT FIGURE 3. Research topics. ABOUT HERE]**

*4.7. Type of software tool*

As discussed above, workflow systems and BPMS have a very similar technology. In fact, they have been considered in some researches as synonyms (Reijers et al., 2016), however it is necessary to know how to distinguish them. From a theoretical point of view, the big difference between both systems is that the workflows need programming and it is not possible to integrate them easily with other software tools. However, BPMS are modelling applications that, when faced with a change in the designed process, do not necessarily need programming, and can be integrated with other existing business applications (Arjonilla Domínguez and Medina Garrido, 2013). Due to the above, it seems interesting to determine which of the analysed works refer to a BPMS tool or to a workflow. In this way it has been found that there are 18 works that analyze the BPMS and there are only four studying workflow systems, they are number [1], [19], [21] and [22] of the appendix (see Figure 4).

**[INSERT FIGURE 4. Type of software tool. ABOUT HERE]**

## 5. Conclusions

There are many business processes within the port activity that can be improved through the use of BPMS, which can help to coordinate and automate critical processes. The implementation of BPMS also offers important connectivity opportunities with other technologies, such as those included in Industry 4.0. The objective of this work is to demonstrate, through a systematic analysis of the literature, the state of the art in BPMS research in the port industry. This systematic literature review has served as a means of evaluating and interpreting all relevant research on this particular research issue. From the results obtained, only 22 works have been found in the academic field that relates BPMS and port industry, which contrasts with the use of these systems in other industries. However, as has been mentioned above, this technology is increasingly being implemented and developed in the port business environment and it is, therefore, not surprising that the research trend in this area is bullish. The eastern countries are the ones that most research this paradigm. They do it from case studies on BPMS rather than workflow, by researching the modeling of processes in ports, and from the area of engineering fundamentally. However, some limitations should be taken into account, specifically, the application of content analysis does not allow us to exempt the study of certain subjectivity. To avoid or minimize this possible bias, two different reviewers, with a third reviewer intervening to solve the discrepancies, have carried out this analysis. Another limitation could be found in the number of databases checked. Perhaps, if the search is extended to a greater number of repositories, we could find more works relating both topics. In any case, there were only include those papers published in academic journals to guarantee their quality.

The main practical implication that this study entails for port management is that the growing use of BPMS has reached the port sector, meaning that port CEOs must consider the value of BPMS for the continuous improvement and flexibility of their processes and to improve the connectivity of technologies within the 4.0 industry. In regard to the theoretical implications of this study, the findings show that there is a lack of studies that relates BPMS and the port sector (practically non-existent), which represents an important gap to be covered and a future line of research. This paper offers researchers the state of the art on BPMS in ports, allows them to know the current and emerging technological trends, points out the most treated topics and which ones present gaps that could be used for future lines of research. As commented before, one of the gaps identified is the incomplete analysis of the stages of the BPM life cycle. While the literature has worked the stages of implementation, modelling and architecture of BPMS in ports, the stages of analysis, execution, control and optimization of processes have not yet been analysed in this sector. On the other hand, the lack of empirical quantitative studies is evidenced. While the interest of the case study in analysing the particularities of each port is evident, studying the success of the implementation of BPMS in ports would require more representative research. In addition, a call could be made for researchers in the business organization area to address these issues, since existing studies come from more technical areas.

## 6. References

Aalst, W. Van Der *et al.* (2003) 'Business process management: A survey', in *International conference on business process management*. Springer Berlin Heidelberg, pp. 1–12.

Alvesson, M. and Sandberg, J. (2011) 'Generating research questions through problematization', *Academy of Management Review*, 36(2), pp. 247–271.

Amadi-echendu, A. and Kruger, L. (2015) 'Supply chain integration in the South African conveyancing environment', *Journal of Transport and Supply Chain Management*, 10(1), pp. 1–13. doi: 10.4102/jtscm.v10i1.211.

Andrade Longaray, A. *et al.* (2015) 'Use of Bpm To Redesign the Container Handling Process: a Brazilian Retroportuary Terminal Case', *Independent Journal of Management & Production*, 6(3), pp. 667–686. doi: 10.14807/ijmp.v6i3.294.

Arjonilla Domínguez, S. J. and Medina Garrido, J. A. (2013) *La gestión de los sistemas de información en la empresa*. 3rd edn, *Colección Economía y empresa/Pirámide*. 3rd edn. Madrid: Ediciones Pirámide.

Bernardino Araújo, M., Rodrigues Filho, B. A. and Franco Gonçalves, R. (2016) 'Business Process Management Notation for a costing model conception', *Brazilian Journal of Operations & Production Management*, 13(3), pp. 244–251.

Bhosale, V. A. and Kant, R. (2016) 'Metadata analysis of knowledge management in supply chain', *Business Process Management Journal*. Bradford: Emerald Group Publishing, Limited, 22(1), pp. 140–172.

Bran, C. (2015) 'The Flexibilization of Information Systems', *FAIMA Business & Management Journal*, 3(4), pp. 64–75.

Britto, P. A. P. De *et al.* (2015) 'Promoção da concorrência no setor portuário: Uma análise a partir dos modelos mundiais e aplicação ao caso brasileiro', *Revista de Administração Pública*, 49(1), pp. 47–71. doi: http://dx.doi.org/10.1590/0034-76121690.

Brocke, J. Vom *et al.* (2014) 'Ten principles of good business process management.', *Business Process Management Journal*. Edited by D. Thomas Kohlborn, Dr Oliver Mueller,. Emerald Group Publishing Limited, 20(4), pp. 530–548. doi: 10.1108/JFM-03-2013-0017.

Castro, A. A. and Erazo, S. C. R. (2009) 'Direccionamiento estratégico apoyado en las TIC', *Estudios Gerenciales*. Santiago de Cali: Universidad Icesi, 25(111), pp. 127–143.

Davenport, T. H. and Short, J. E. (1990) 'The New Industrial Engineering: Information Technology And Business Process Redesign', *Sloan management review*. Cambridge, United States, Cambridge: Massachusetts Institute of Technology, 31(4), pp. 11–27.

Dumas, M. *et al.* (2013) 'Fast detection of exact clones in business process model repositories', *Information Systems*, 38(4), pp. 619–633. doi: 10.1016/j.is.2012.07.002.

Dybå, T. and Dingsøyr, T. (2008) 'Empirical studies of agile software development: A systematic review', *Information and Software Technology*, 50(9–10), pp. 833–859. doi: 10.1016/j.infsof.2008.01.006.

Echeverri, D. R. C. and Cruz, R. Z. (2014) 'Revisión de instrumentos de evaluación de clima organizacional/Review of organizational climate assessment tools/Análise de instrumentos de avaliação de clima organizacional', *Estudios Gerenciales*. Santiago de Cali: Universidad Icesi, 30(131), pp. 184–189.

Feng, Z., Ye, Y. and Liu, X. (2011) 'A formal modeling method for grid workflow based on concurrent transaction logic', *Journal of Convergence Information Technology*. College of Zhijiang, Zhejiang University of Technology, Hangzhou 310024, China, 6(2), pp. 59–69. doi: 10.4156/jcit.vol6.issue2.7.

Ferretti, M. and Schiavone, F. (2016) 'Internet of Things and business processes redesign in seaports: The case of Hamburg', *Business Process Management Journal*. Bradford: Emerald Group Publishing, Limited, 22(2), pp. 271–284.

Fioresi De Sousa, E. *et al.* (2019) 'Port Worker and Operation in Ship'S Hold: a Study in the Port Environment of the Espírito Santo State', *Revista Produção Online*, 19(2), pp. 430–448.

Friedman, T. L. (2005) 'It's a flat world, after all', *The New York Times*, 3, pp. 33–37.

Guinea, A. S., Nain, G. and Le Traon, Y. (2016) 'A systematic review on the engineering of software for ubiquitous systems', *The Journal of Systems and Software*. New York: Elsevier Sequoia S.A., 118, p. 251.

Hammer, M. and Champy, J. (1993) 'Reengineering the corporation: A manifesto for business revolution', *Business horizons*. Elsevier, 36(5), pp. 90–91.

Heilig, L., Lalla-Ruiz, E. and Voß, S. (2017) 'Digital transformation in maritime ports : analysis and a game theoretic framework', *Netnomics*. NETNOMICS: Economic Research and Electronic Networking, 18(2–3), pp. 227–254.

Hitpass, B. and Astudillo, H. (2019) 'Editorial : Industry 4 . 0 Challenges for Business Process Management and Electronic-Commerce BPM as Support for Industry 4 . 0 and E- Commerce', *Journal of Theoretical and Applied Electronic Commerce Research*, 14(1), pp. I–III. doi: 10.4067/S0718-18762019000100101.

Idri, A., Hosni, M. and Abran, A. (2016) 'Systematic literature review of ensemble effort estimation', *The Journal of Systems and Software*. New York: Elsevier Sequoia S.A., 118, pp. 151–175.

Işik, Ö. *et al.* (2013) 'Practices of knowledge intensive process management: Quantitative insights', *Business Process Management Journal*. Bradford: Emerald Group Publishing, Limited, 19(3), pp. 515–534. doi: 10.1108/14637151311319932.

Jalali, S. and Wohlin, C. (2012) 'Systematic literature studies: database searches vs. backward snowballing', *Proceedings of the ACM-IEEE international*, pp. 29–38.

Kitchenham, B. and Brereton, P. (2013) 'A systematic review of systematic review process research in software engineering', *Information and Software Technology*, 55(12), pp. 2049–2075. doi: 10.1016/j.infsof.2013.07.010.

Kitchenham, B. and Charters, S. (2007) 'Guidelines for performing Systematic Literature reviews in Software Engineering Version 2.3', *Engineering*, 45(4ve), p. 1051. doi: 10.1145/1134285.1134500.

Kohlbacher, M. (2010) 'The effects of process orientation: a literature review', *Business Process Management Journal*. Bradford: Emerald Group Publishing, Limited, 16(1), pp. 135–152. doi: http://dx.doi.org/10.1108/14637151011017985.

Kravchenko, A. V., & Makarenko, M. V. (2017) 'Directions in port management improvement basing on the methods of correlation and regression analysis', *Aktual'ni Problemy Ekonomiky= Actual Problems in Economics*, 189, pp. 325–330.

Malaurent, J. and Avison, D. (2016) 'Reconciling global and local needs: A canonical action research project to deal with workarounds', *Information Systems Journal*, 26(3), pp. 227–257. doi: 10.1111/isj.12074.

Martín-Navarro, A., Lechuga Sancho, M. P. and Medina-Garrido, J. A. (2018) 'BPMS for management: A systematic literature review', *Revista Española de Documentacion Cientifica*, 41(3). doi: 10.3989/redc.2018.3.1532.

Martín, A. (2017) *El impacto de los BPMS en la gestión de los procesos y del conocimiento de las organizaciones*. Universidad de Cádiz (España).

De Martino, M., Carbone, V. and Morvillo, A. (2015) 'Value creation in the port: opening the boundaries to the market', *Maritime Policy & Management*. Routledge, 42(7), pp. 682–698. doi: 10.1080/03088839.2015.1078010.

Metalla, O., Koxhaj, A. and Vyshka, E. (2015) 'Container Terminal and Associating Infrastructure', *Interdisplinary Journal of Research and Development*, II(1).

Minonne, C. and Turner, G. (2012) 'Business Process Management-Are You Ready for the Future?', *Knowledge and Process Management*, 19(3), pp. 111–120. doi: 10.1002/kpm.1388.

Molina Serrano, B. *et al.* (2018) 'Use of Bayesian Networks to Analyze Port Variables', *Logistics*, 2(1), p. 5. doi: 10.3390/logistics2010005.

Paschek, D., Ivascu, L. and Draghici, A. (2018) 'Knowledge Management – The Foundation for a Successful Business Process Management', *Procedia - Social and Behavioral Sciences*. The Author(s), 238, pp. 182–191. doi: 10.1016/j.sbspro.2018.03.022.

Pistol, L. and Bucea-Manea-Tonis, R. (2015) 'Workflow systems as a tool for small and medium size enterprises business processes management', *Journal of Applied Economic Sciences*, 10(8), pp. 1250–1258.

Ramírez-Nafarrate, A. *et al.* (2017) 'Impact on yard efficiency of a truck appointment system for a port terminal', *Annals of Operations Research*. Springer US, 258(2), pp. 195–216. doi: 10.1007/s10479-016-2384-0.

Ramírez Correa, P. and García Cruz, R. (2005) 'Meta- análisis sobre la implantación de sistemas de planificación de recursos empresariales (ERP)', *JISTEM - Journal of Information Systems and Technology Management*, 2(3), pp. 245–273.

Reijers, H. A. (2006) 'Implementing BPM systems: the role of process orientation', *Business Process Management Journal*. Bradford: Emerald Group Publishing, Limited, 12(4), p. 389. doi: http://dx.doi.org/10.1108/14637150610678041.

Reijers, H. A. A., Vanderfeesten, I. and van der Aalst, W. M. P. M. P. (2016) 'The effectiveness of workflow management systems: A longitudinal study.', *International Journal of Information Management*, 36(1), pp. 126–141. doi: 10.1016/j.ijinfomgt.2015.08.003.

Rhee, S.-H., Bae, H. and Choi, Y. (2007) 'Enhancing the efficiency of supply chain processes through web services', *Information Systems Frontiers*, 9(1), pp. 103–118. doi: 10.1007/s10796-006-9020-5.

Rocha, R. dos S. *et al.* (2013) 'The use of software product lines for business process management: A systematic literature review', *Information and Software Technology*. Amsterdam: Elsevier Science Ltd., 55(8), pp. 1355–1373. doi: 10.1016/j.infsof.2013.02.007.

Roeser, T. and Kern, E.-M. (2015) 'Surveys in business process management - a literature review', *Business Process Management Journal*. Bradford: Emerald Group Publishing, Limited, 21(3), pp. 692–718.

Rooney, A., Boyles, A. and Wolfe, M. (2014) 'Systematic review and evidence integration for literature-based environmental health science assessments', *Environmental*, 7(122), pp. 711–722.

Ruiz-Torres, A. J. *et al.* (2018) 'Logistic services in the Caribbean region: An analysis of collaboration, innovation capabilities and process improvement', *Academia Revista Latinoamericana de Administracion*, 31(3), pp. 534–552. doi: 10.1108/ARLA-03-2017-0078.

Smith, H. and Fingar, P. (2003) *Business process management :the third wave*. Tampa, Fla.: Meghan-Kiffer Press.

Tarhan, A., Turetken, O. and Reijers, H. A. (2016) 'Business process maturity models: A systematic literature review', *Information and Software Technology*. Amsterdam: Elsevier Science Ltd., 75, pp. 122–134. doi: 10.1016/j.infsof.2016.01.010.

Tucek, D. (2015) 'The Main Reasons for Implementing BPM in Czech Companies', *Journal of Competitiveness*. Zlin: Tomas Bata University in Zlin, Faculty of Management and Economics, 7(3). doi: http://dx.doi.org/10.7441/joc.2015.03.09.

Turner, M. *et al.* (2010) 'Does the technology acceptance model predict actual use? A systematic literature review', *TAIC-PART 2008 - TAIC-PART 2008*, 52(5), pp. 463–479. doi: DOI: 10.1016/j.infsof.2009.11.005.

Unterkalmsteiner, M. *et al.* (2012) 'Evaluation and Measurement of Software Process Improvement—A Systematic Literature Review', *IEEE Transactions on Software Engineering*, 38(2), pp. 398–424.

Wen, J. *et al.* (2012) 'Systematic literature review of machine learning based software development effort estimation models', *Information and Software Technology*, 54, pp. 41–59. doi: 10.1016/j.infsof.2011.09.002.

Wong, W. P. (2013) 'Business-process management: a proposed framework for future research', *Total Quality Management & Business Excellence*. Abingdon: Taylor & Francis Ltd., 24(5–6), pp. 719–732. doi: 10.1080/14783363.2013.776773.

Yeo, G.-T., Roe, M. and Dinwoodie, J. (2011) 'Measuring the competitiveness of container ports: Logisticians' perspectives', *European Journal of Marketing*, 45(3), pp. 455–470. doi: http://dx.doi.org/10.1108/03090561111107276.

## *Appendix*

[1] Bassil S., Keller R.K., Kropf P. (2004) A Workflow-Oriented System Architecture for the Management of Container Transportation. In: Desel J., Pernici B., Weske M. (eds) Business Process Management. BPM 2004. Lecture Notes in Computer Science, vol 3080. Springer, Berlin, Heidelberg.

[2] Besri, Z., & Boulmakoul, A. (2017). Framework for organizational structure re-design by assessing logistics' business processes in harbor container terminals. *Transportation research procedia*, *22*, 164-173.

[3] Cai, H., Hu, Z. H., & Liu, W. (2009, November). Container process flow visualization by spatial location technologies. In *Computational Intelligence and Industrial Applications, 2009. PACIIA 2009. Asia-Pacific Conference on* (Vol. 1, pp. 131-134). IEEE.

[4] Cimino, M. G., Palumbo, F., Vaglini, G., Ferro, E., Celandroni, N., & La Rosa, D. (2017). Evaluating the impact of smart technologies on harbor's logistics via BPMN modeling and simulation. *Information Technology and Management*, *18*(3), 223-239.

[5] Fady, R. I. (2013, January). Evaluate the E-business Maturity of the Port of Alexandria. In *System Sciences (HICSS), 2013 46th Hawaii International Conference on* (pp. 3888-3899). IEEE.

[6] Fady, R. I. and Beeson, I., "Drawing out the Essential Business of Ports", 11[th] IBIMA Proceedings, Cairo, Egypt, 2009.

[7] Froese, J. (2007). Seaport Integration and Networking–A European Case Study–. *IFAC Proceedings Volumes*, *40*(19), 124-129.

[8] Geldof, F., Van Haarlem, B. C., Lock, W., & Roubtsova, E. E. (2008, April). Modelling and simulation of high capacity waterside container handling systems at deep-sea terminals. In *Proceedings of the 2008 Spring simulation multiconference* (pp. 198-205). Society for Computer Simulation International.

[9] Islam, S., Olsen, T., & Daud Ahmed, M. (2013). Reengineering the seaport container truck hauling process: Reducing empty slot trips for transport capacity improvement. *Business process management journal*, *19*(5), 752-782.

[10] Khalifa, I. H., El Kamel, A., & Yim, P. (2011). Transportation process of containers BPMN-modelling and transformation into ACTIF model. *ROMJIST*, *14*(1), 67-80.

[11] Longaray, A. A., Munhoz, P. R., Albino, A. S., & Castelli, T. M. (2015). Use of BPM to redesign the container handling process: a brazilian retroportuary terminal case. *Independent Journal of Management & Production*, *6*(3), 667-686.

[12] Lyridis, D. V., Fyrvik, T., Kapetanis, G. N., Ventikos, N., Anaxagorou, P., Uthaug, E., & Psaraftis, H. N. (2005). Optimizing shipping company operations using business process modelling. *Maritime Policy & Management*, *32*(4), 403-420.

[13] Nisafani, A. S., Park, J., Bae, H., & Yahya, B. N. (2012). Container Flow Management in Port Logistics Based on BPM Framework. Journal of Information Technology and Architecture, 9(1), 1-10.

[14] Pulshashi, I. R., Bae, H., Sutrisnowati, R. A., Yahya, B. N., & Park, S. (2015). Slice and Connect: Tri-Dimensional Process Discovery with Case Study of Port Logistics Process. *Procedia Computer Science*, *72*, 461-468.

[15] Spagnolo, G. O., Marchetti, E., Coco, A., Scarpellini, P., Querci, A., Fabbrini, F., & Gnesi, S. (2016, January). An experience on applying process mining techniques to the Tuscan port community system. In *International Conference on Software Quality* (pp. 49-60). Springer, Cham.

[16] Stevanov, B., Zuber, N., Šostakov, R., Tešić, Z., Bojić, S., Georgijević, M., & Zelić, A. (2016). Reengineering the Port Equipment Maintenance Process. *International Journal of Industrial Engineering and Management*, *7*(3), 103-109.

[17] Świeboda, J. (2016). Analysis and assessment of an information subsystem in an inland container terminal/Analiza i ocena podsystemu informacyjnego w lądowym terminalu kontenerowym. *Journal of KONBiN*, *38*(1), 99-130.

[18] Wang, Y., Caron, F., Vanthienen, J., Huang, L., & Guo, Y. (2014). Acquiring logistics process intelligence: Methodology and an application for a Chinese bulk port. *Expert Systems with Applications*, *41*(1), 195-209.

[19] Wang, Y., Huang, L., & Guo, Y. (2013). Integrating declarative and imperative approach to model logistics service processes. *Journal of Industrial Engineering and Management*, *6*(1), 237-248.

[20] Yahya, B. N., Mo, J., Bae, H., & Lee, H. (2010, July). Ontology-based process design support tool for vessel clearance system. In Computers and Industrial Engineering (CIE), 2010 40th International Conference on (pp. 1-6). IEEE.

[21] Yang, B., & Liu, L. (2007). A Case Study of Enterprise Application Integration Based on Workflow Management System. In Research and Practical Issues of Enterprise Information Systems II (pp. 425-431). Springer, Boston, MA.

[22] Zhang, H., Collart-Dutilleul, S., & Mesghouni, K. (2013). Seaport Productivity Research basing on P-time Petri Net. IFAC Proceedings Volumes, 46(24), 101-106.

# TABLES

**Table I. Quality Criteria**

| No. | Pregunta | Autor |
| --- | --- | --- |
| QA1 | Are we faced with an empirical study? | Dybå y Dingsøyr, 2007 |
| QA2 | Are the objectives of the study (or review) clearly defined? | Dybå y Dingsøyr, 2007; Kitchenham and Bereton, 2013 |
| QA3 | Is there adequate description of the context in which the study was carried out? | Dybå y Dingsøyr, 2007; Kitchenham and Bereton, 2013 |
| QA4 | Was the research method or methodology appropriate to address the objectives of the study? | Kitchenham and Bereton, 2013 |
| QA5 | Was the data analysis sufficiently rigorous? | Kitchenham and Bereton, 2013 |

| QA6 | Are the results of the evaluation (study) clearly defined? | (Unterkalmsteiner *et al.*, 2012); Kitchenham and Bereton, 2013 |
| QA7 | Are the limitations of the study explicitly analyzed? | Wen et al., 2012 |
| QA8 | Is the study of value to the scientific community and business community? | Wen et al., 2012; Kitchenham and Bereton, 2013 |

Source: Adapted from Dybå & Dingsøyr (2008)

## Table II. Contributions by country.

| Country | Articles | % |
|---|---|---|
| China [3][18][19][21] | 4 | 18.18 |
| South Korea [13][14][20] | 3 | 13.64 |
| France [2][22] | 2 | 9.09 |
| Italy [4][15] | 2 | 9.09 |
| Germany [7] | 1 | 4.55 |
| Brazil [12] | 1 | 4.55 |
| Canada [1] | 1 | 4.55 |
| Egypt 6] | 1 | 4.55 |
| Greece [13] | 1 | 4.55 |
| Holland [8] | 1 | 4.55 |
| Kuwait [5] | 1 | 4.55 |
| Morocco [2] | 1 | 4.55 |
| New Zealand [9] | 1 | 4.55 |
| Poland [16] | 1 | 4.55 |
| Serbia [17] | 1 | 4.55 |

## Table III. Contributions by area of knowledge

| Area of knowledge | Articles | % |
|---|---|---|
| Engineering [1][4][8][14][15][16][17][20] | 8 | 36.36 |
| Computer [2][10][12][22] | 4 | 18.18 |
| Business Administration [3][11][18][19] | 4 | 18.18 |
| Information System [5][6][9][21] | 4 | 18.18 |
| Logistic [7][13] | 2 | 9.09 |

## Tabla IIV. Type of study

| Tipo de estudio | Articles | % |
|---|---|---|
| Case study [2][6][9][10][11][12][13][14][15][16][17][18][19][21] | 14 | 63.64 |
| Theoretical [1][3][5][7][8][20][22] | 7 | 31.82 |
| Experimental [4] | 1 | 4.55 |

## FIGURES

**Figure 1. Phases of the systematic literature review process and number of definitive articles**

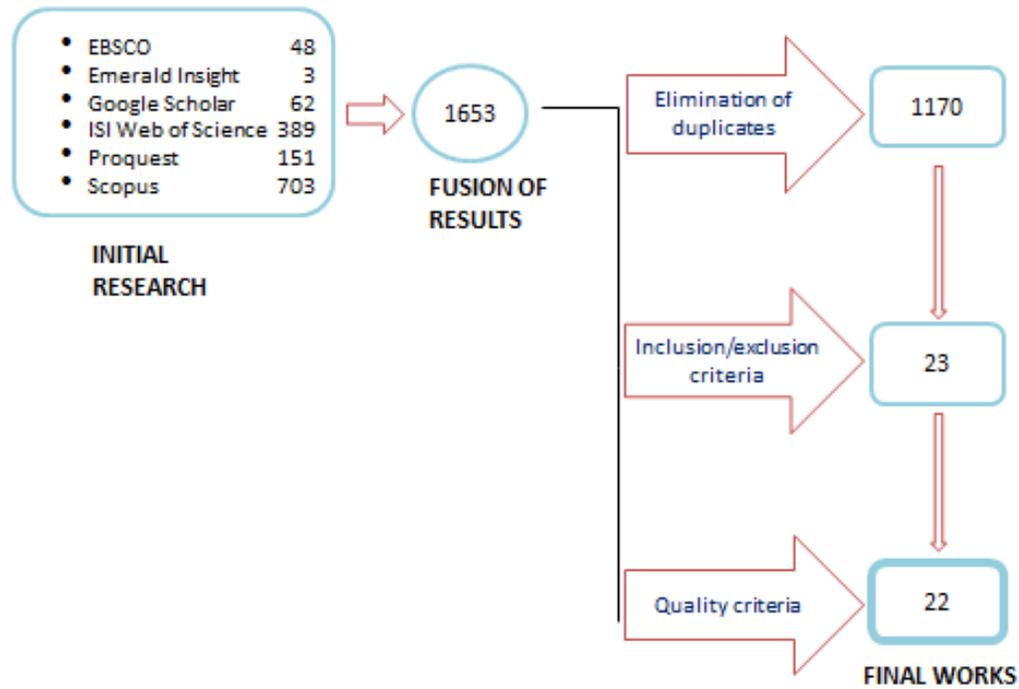

**Figure 2. Number of publications per year.**

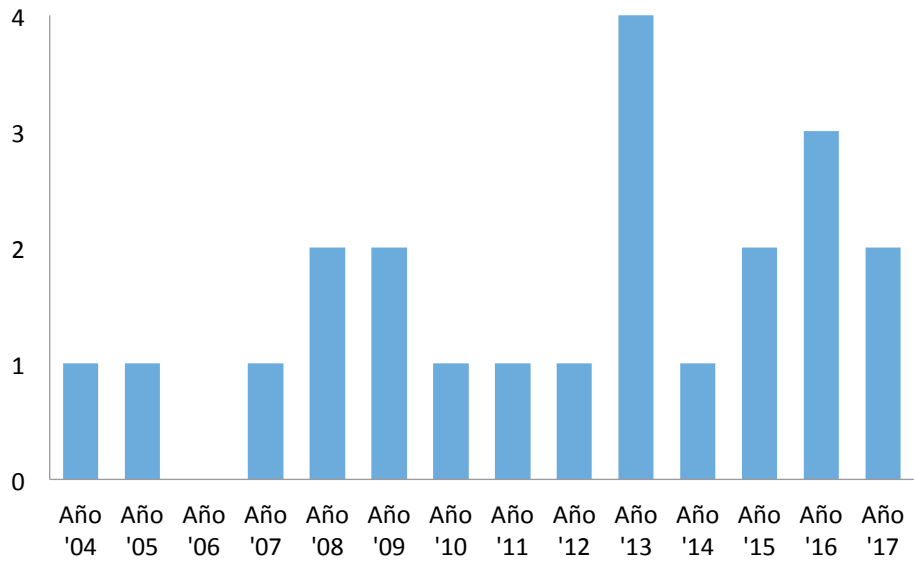

**Figure 3. Research topics**

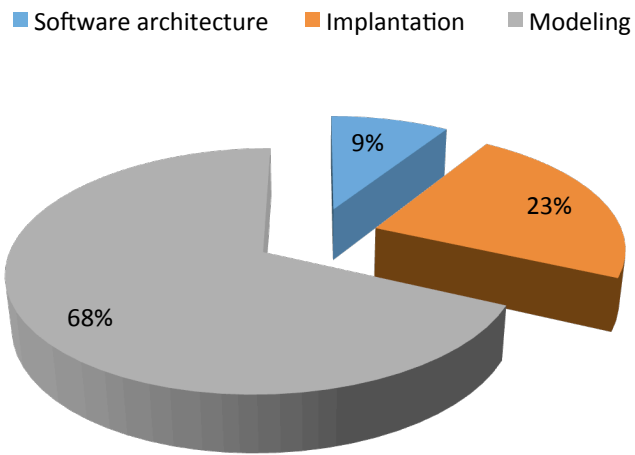

**Figure 4. Type of software tool**

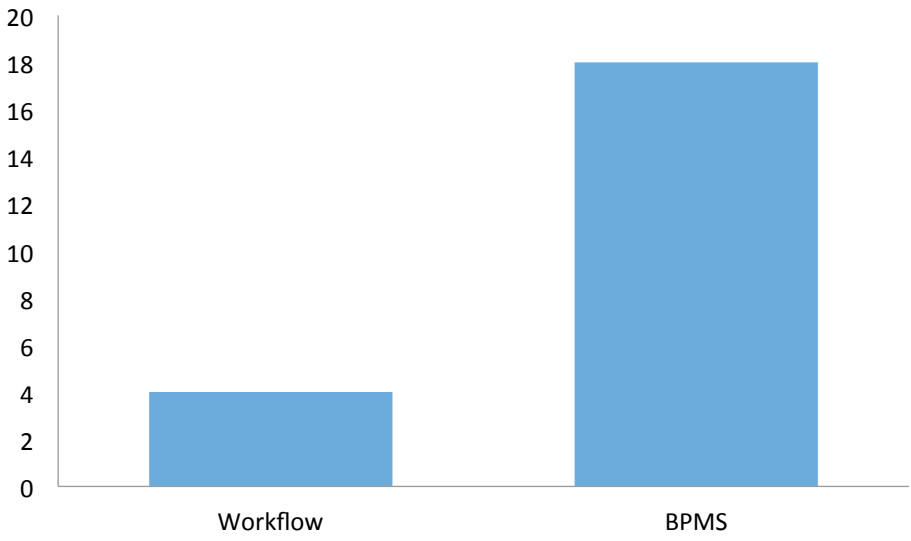